# Large-gap quantum spin Hall state in functionalized dumbbell stanene

Ya-ping Wang,[a] Wei-xiao Ji,[a] Chang-wen Zhang*,[a] Ping Li,[a] Feng Li,[a] Pei-ji Wang,[a] Sheng-shi Li,[b] Shi-shen Yan [b]

[a] *School of Physics and Technology, University of Jinan, Jinan, Shandong, 250022, People's Republic of China*

[b] *School of Physics, State Key laboratory of Crystal Materials, Shandong University, Jinan, Shandong, 250100, People's Republic of China*

Two-dimensional dumbbell (*DB*) stanene has been proposed as a promising candidate material for realizing quantum spin Hall effect (QSHE) by Tang *et al* [P. Tang, P. Chen, W. Cao, H. Huang, S. Cahangirov, L. Xian, Y. Xu, S. C. Zhang, W. Duan, A. Rubio. *Phys. Rev. B*, 90, 121408 (2014)]. However, the small bulk-gap limits its possible applications at room temperature. Based on first-principles calculations, we predict that its band gap can be enhanced to 148 meV under methyl-functionalization, which can be further tuned by applying lattice strain. The QSHE is confirmed by $s$-$p_{x,y}$ band inversion, topological invariant $Z_2 = 1$, and helical gapless edge within bulk band gap. Notably, the characteristic properties of edge states, such as the large Fermi velocity and Dirac cone, can be modulated by edge modification. The effects of substrates on topological properties are explored when it is grown on various substrates, like SiC, *h*-BN, and $Bi_2Te_3$ sheets. These findings provide significant guidance for future fabrication and realistic applications of QSHE based on stanene in spintronics.

*** Corresponding author:** C. W. Zhang: zhchwsd@163.com TEL: 86-531-82765976

Two-dimensional (2D) topological insulators (TIs), known as quantum spin Hall insulators (QSHIs), first predicted by Kane and Mele [1, 2] have sparked off extensive research to develop this new class of materials. The 2D QSHIs are characterized by insulating bulk states with gapless edge states [3–5]. Several quantum well structures [6–9] have been firstly reported to be QSHIs, but their bulk-gaps are not large enough for wide practical applications. In the Kane-Mele model of graphene [1], the bulk-gap of a QSHI was determined primarily by spin-orbit coupling (SOC). However, due to the rather weak SOC, the associated gap is tiny (~$10^{-3}$ meV), where the quantum spin Hall effect (QSHE) can only be observed at an unrealistically low temperature. Based on this point, some new 2D materials, which are composed of heavy elements [10–16], have been predicted as the candidates for large gap QSHIs. However, the observations of QSHE are still challengeable experimentally.

Tin (Sn) atom, the counterpart of carbon among group-IV honeycomb lattice, is known for its stronger SOC which can drive nontrivial TI states. It has been gained great interest, due to unique structural and electronic properties, as well as its better compatibility with current Si nanotechnology [17-22]. Stanene is a QSHI with a band gap of ~ 0.1 eV induced by strong SOC [17]. Such an upper limit, however, can be significantly broken through and enhanced to 0.3 eV once functionalized with hydrogen or halogen atoms [17, 19]. In addition, by single-side functionalization, a near room-temperature quantum anomalous Hall effect (QAHE) can be realized instead [18]. Experimentally, ultrathin stanene film has been recently fabricated by molecular beam epitaxy (MBE) on a $Bi_2Te_3$ (111) substrate [21], which indeed demonstrates the existence of stanene. But the valence bands of stanene are pinned with conduction bands of $Bi_2Te_3$ (111), giving metallic states. In sharp contrast to graphene [1] and silicene, [23, 24] the bond lengths in stanene are much larger, therefore the relatively weak π-π bonding cannot stabilize hexagonal structure, resulting in instability of stanene. [25] A dumbbell (*DB*) stanene, containing some fourfold-coordinated tin atoms, is proposed to be a stable phase. [26] *DB* stanene is a 2D QSHI with an inverted $p_{x,y}$-$p_z$ band order, but its band gap is rather smaller. In addition, about 40% threefold-coordinated Sn atoms still exist, which may be chemically reactive and thus to the disadvantage of practical applications. To solve this problem, Chen *et al* [25] have predicted a hydrogenated *DB-Sn* as a large-gap 2D TI. However, the experiment [27] reveals that the plasma hydrogenation generally exhibit quick kinetics, with rapid increase of defects and lattice disorder even under short plasma exposures. Also, the hydrogen-functionalized systems are not stable in a long term and are greatly reactive in air, as there is extreme propensity to oxidize under ambient condition. Thus, the achievement of hydrogenated systems of high quality is rather challengeable, demonstrating that the observation of QSHE is difficult in experiments.

In this work, we propose to saturate these dangling bonds on three-coordinated Sn(2) atoms with organic molecule methyl ($CH_3$), named as *DB*-$SnCH_3$ [see Fig. 1(d)], motivated by the recent experiment [28]. Based on first-principles calculations, we predict that the *DB*-$SnCH_3$ is stable and behaves as a 2D TI, with a bulk-gap as large as 148 meV. Notably, the characteristic properties of edge states, such as the high Fermi velocity and Dirac cone, can be tuned by edge modifications. We also propose SiC, *h*-BN, and $Bi_2Te_3$ as appropriate substrates to support *DB*-$SnCH_3$ preserving the large-gap QSHE. As an ideal platform for QSHE applications, some exotic phenomena, such as QAHE [29-31] and Chern half-metallicity [32], may be realized in these structures.

First-principles calculations are performed by using density functional theory (DFT) [33] methods as implemented in the Vienna ab initio simulation package (VASP) [34]. The projector-augmented-wave (PAW) potential [35,36], Perdew-Burke-Ernzerhof (PBE) exchange-correlation functional [37], and the plane-wave basis with a kinetic energy cutoff of 500 eV are employed. The Brillouin zone is sampled by using a $9\times9\times1$ Gamma-centered Monkhorst–Pack grid. The vacuum space is set to 20 Å to minimize artificial interactions between neighboring slabs. During the structural optimization of *DB*-$SnCH_3$, all atomic positions and lattice parameters are fully relaxed, and the maximum force allowed on each atom is less than 0.02 eV/ Å. SOC is included by a second variational procedure on a fully self-consistent basis.

*DB* stanene has a hexagonal structure with space group $D_{6h}$ (P6/mmm) and ten Sn atoms in one unit cell. [26] In this structure, the Sn (1) staying in the same plane is fourfold-coordinated, while Sn (2) at up and down positions is threefold-coordinated with one dangling bond, see Fig. 1(a). The height of DB stanene ($h$) and the distance between neighboring *DB* stanene ($a = a_1 = a_2$) are 3.42 and 9.05 Å, respectively, which is consistent with the results in Ref. [26]. The calculated band structures of *DB* stanene are shown in Figs. 1(b) and 1(c). An indirect semiconducting character is obtained without SOC, while a $p_{x,y}$-$p_z$ band inversion at Γ point occurs when the SOC is turned on. Thus *DB* stanene is a QSHI, with a small band gap of 22.4 meV. Such an upper limit, however, can be significantly broken through, as discussed in the following.

Chemical functionalization of 2D materials is an efficient way to create new QSHIs with desirable features, as in other hexagonal structures [17, 18, 20]. In this work, the threefold-coordinated Sn (2) atom is decorated by organic molecule methyl ($CH_3$), named as *DB*-$SnCH_3$, see Fig. 1(d). After structural optimization, the buckling height *h* between Sn (2) atoms increases by 0.08 Å, along with lattice constant decreasing by 0.05 Å, as compared with those of *DB* stanene. In addition, the calculated formation energy of *DB*-$SnCH_3$ turns out to be -17.6 meV/Å$^2$, larger than that of $GeCH_3$ (-5.7 meV/Å$^2$) [38]. This indicates that the methyl can strongly bind to Sn(2) atoms by a chemical bond, showing a higher thermodynamic stability relative to their elemental reservoirs. The phonon spectrum along highly symmetric directions is also calculated, as is shown in Fig. S1 in supplementary information [39]. No imaginary phonon mode is found here, confirming the stability of *DB*-$SnCH_3$ film.

The band structures of *DB*-$SnCH_3$ are displayed in Figs. 1(e) (f). In the absence of SOC, it is gapless with the valence and conduction bands meeting at the Fermi level. The valence band maximum (VBM) and conduction band minimum (CBM) arise mainly from the $p_{x,y}$ orbitals of Sn(1) atoms, which are energetically degenerate at the Γ point as a direct result of $C_{3v}$ lattice symmetry. The valence bands nearest to VBM, however, are mainly from the *s* and $p_z$ orbitals of Sn(1) and Sn(2), with $p_z$ bands mainly located at the K point, see Fig. 1(e). Such band alignment has also been reported for the functionalized 2D stanene [17, 18, 20]. When taking SOC into account, a band gap of 327 meV is opened up at the Γ point, along with an indirect-gap of 148 meV [Fig. 1(f)], which is much larger than the intrinsic DB stanene, see Fig. 1(c). As observed in previously reported 2D TIs like $ZrTe_5$, $HfTe_5$ [40], and GaSe, [41] the SOC induced band gap opening at the Fermi level is a strong indication of the existence of topologically nontrivial phase. To confirm band topology in 2D *DB*-$SnCH_3$, we calculate the topologically invariant $Z_2$ following the approach proposed by Fu and Kane [42]. As illustrated in Fig. 2(b), the products of the parity eigenvalues at three symmetry points: M (0.0, 0.5), M (0.5, 0.5) and M (0.5, 0.0) are both -1, while at the Γ (0.0, 0.0) displays +1, yielding $Z_2 = 1$. We also check the influence of randomness and coverage to the topological properties, see Fig. S2 in supplementary information.[39]

The most important performance for TIs is the existence of helical edge states with spin-polarization protected by TRS, which can be calculated with the help of Wannier 90 package [43]. Based on maximally localized Wannier functions (MLWFs) of *p* orbitals, the edge Green's function [44] of the semi-infinite lattice is constructed using the recursive method, and the local density of state (LDOS) is presented in Fig. 2(c). Obviously, the edge bands connect completely the conduction and valence bands and span the 2D bulk band gap, yielding a 1D gapless edge states. By identifying the spin-up (↑) and spin-down (↓) contributions in the edge spectral function, the counter-propagating edge states can exhibit opposite spin-polarization, as shown in Fig. 2(d), in accordance with the spin-momentum locking of 1D helical electrons. Regrettably, an additional dangling bond state within the bulk gap occurs, unfavorable for spin-polarized transport currents.

To gain insight into the origin of the dangling bond state, we further perform tight binding (TB) calculations on the zigzag nanoribbon, with edge Sn atoms terminated by hydrogen atoms,[45] see Fig. 2(f). The calculated band structures are displayed in Fig. 2(e), illustrating the remarkably different topological edge states under different chemical edge environments. (*i*) The edge states in hydrogenated case contain a single Dirac cone with its node lying in the middle of the bulk gap, while the node of the Dirac cone without H adsorption lies above the CBM; (*ii*) The original extended gapless edge states in Fig. 2(c) become localized in the k-space, similar to the edge states in HgTe/CdTe quantum well. [6, 7] Consequently, the Fermi velocity of Dirac states, as reflected by the band slope, can be enhanced significantly with chemical edge modification. Here, the Fermi velocity at Dirac point located at the band-gap reaches ~ $4.0\times10^5$ m/s, comparable to that of $5.3 \times 10^5$ m/s in HgTe/CdTe quantum well [6,7]. (*iii*) More surprisingly, the bands corresponding to dangling bond states are removed completely, see Fig. 2(e), which means that the energy scattering reduces significantly in process of device applications.

To reveal TI nature, we analyze the band evolution at the Γ point for *DB*-$SnCH_3$, as illustrated Fig. 3(a). The chemical bonding between Sn-Sn atoms make the *s* and $p_{x,y}$ orbitals split into the bonding and anti-bonding states, i.e., $|s^{\pm}>$ and $|p_{x,y}^{\pm}>$, where the superscripts + and − represent the parities of corresponding states, respectively. In contrast to *DB* stanene, see Figs. 1(b) and (c), in which the band inversion originates from $|p_{x,y}^{+}>$ and $|p_z^{-}>$, the methyl functionalization makes $|p_z^{-}>$ increase in energy, thus $|s^{-}>$ and $|p_{x,y}^{+}>$ locate near the Fermi level, but both $|s^{-}>$ and $|p_{x,y}^{+}>$ origin from Sn(1), with $|p_{x,y}^{+}>$ being above $|s^{-}>$, see Fig. S3. [39] In this case, $|s^{-}>$ is unoccupied, while the degenerate $|p_{x,y}^{+}>$ is half occupied, resulting in a gapless character, as listed in stage (I). With the inclusion of SOC, the degenerated $|p_{x,y}^{+}>$ splits into $|p, \pm 3/2>$ with a total angular momentum $j = 3/2$ and $|p, \pm 1/2>$ with $j = 1/2$, opening a full band gap (stage II). Thus, it is the methyl functionalization that induces band inversion rather than the SOC which only produces a large band-gap, i.e., the methyl functionalization is responsible for nontrivial band orders.

Since the substrates are inevitable in device applications, a free-standing *DB*-$SnCH_3$ film must eventually be deposited on a substrate. Actually, the high-quality group-IV films have been successfully grown on different substrates based on MBE method. [46] It is reasonable to expect that it can also be transferred on substrate using the similar technique. Considering compatibility of lattice constant, we explore strain-driven TI phase transition in *DB*-$SnCH_3$. Figure 3(b) presents the variation of band gap ($E_g$, $E_\Gamma$) as a function of external strain. Interestingly, the QSHE is sensitive to strain, upon which the band gap of TI and topological phase transitions can be tuned efficiently. Typically, over an achievable range of strains ($\varepsilon$ = -0.03 to 0.06), *DB*-$SnCH_3$ is nontrivial. When the lattice is compressed from $\varepsilon$ = -0.09 to -0.03, the trivial band order occurs, forming a normal insulator (NI). If we further compress strains beyond critical point ($\varepsilon$ < -0.09), the $s$-$p_{x,y}$ inversion occurs again, and thus *DB*-$SnCH_3$ reverts back to QSHI with a gap smaller than one at equilibrium state. On the other hand, when the lattice constant stretches beyond 0.06, the metallic (M) states are obtained. With this in mind, several semiconductor/insulating substrates, such as SiC, *h*-BN, and $Bi_2Te_3$, are selected to construct heterostructures (HTSs) to check whether its nontrivially topological properties can be maintained.

SiC is an important semiconductor to support 2D materials to realize QSHE, including Bi bilayers [47]. Here, we construct two kinds of HTSs, such as monolayer (ML) SiC and 5-ML SiC layers, to support *DB*-$SnCH_3$. Figure 4(a) displays the *DB*-$SnCH_3$@3×3-ML-SiC HTS, where the lattice mismatch between *DB*-$SnCH_3$ and 1ML SiC is only 3.14%, with the distances between adjacent layers being 3.05 Å. For 5-ML SiC supported *DB*-$SnCH_3$, the most stable configuration is shown in Fig. 4(c), in which the two bottom Sn atoms of *DB*-stanene are chemically bonded to Si atoms in the surface SiC layer. In this case, the $p_z$ orbital of Sn(2) atoms are fully decorated by dangling bond of Si atoms, while the remaining Si atoms on surface of 5-ML SiC film are saturated by hydrogen atoms. The calculated formation energies indicate that the interactions between *DB*-$SnCH_3$ and 5-ML SiC strong chemically bonding with an energy of -4.96 eV/per atom, while the former case is a weak van der Waals (vdW) HTS (-0.080 eV/per atom). Figs. 4(b) and (d) display the corresponding band structures in two cases, with the indirect-gap as large as 87.23 and 95.85 meV for ML and 5-ML films, respectively, exceeding the room temperature thermal energy. The $Z_2$ invariant is calculated to identify the nontrivial phase. Note that the band gap of *DB*-$SnCH_3$ decreases clearly (from 148 to 87.23 or 95.85 meV) when it is placed on SiC substrate. The relatively smaller band gap of *DB*-stanene obtained on SiC can be attributed to the lattice-strain induced by the substrate and the strength of effective SOC. These all indicate the proper choice on the substrate is a key factor for maintaining TI gap of free-standing *DB*-stanene.

The *h*-BN sheet, a large band gap insulator with a high dielectric constant, fills the requirement to assemble 2D stacked nano-electronic devices [48]. To check this idea, we construct a *DB*-$SnCH_3$@$\sqrt{13}\times\sqrt{13}$ *h*-BN HTS, see Fig. 5(a), where the lattice mismatch is only 0.53%. After a full relaxation, the distance between adjacent layers is 2.80 Å, with a binding energy of -0.086 eV/atom, indicating that *DB*-$SnCH_3$ interacts weakly with *h*-BN substrate. The calculated band structure with SOC is shown in Fig. 5(b). In this weakly coupled system, the *DB*-$SnCH_3$@*h*-BN remains semiconducting with an inverted band order, and there is essentially no charge transfer between the adjacent layers. If we compare the bands of *DB*-$SnCH_3$ supported with and without *h*-BN, little difference is observed. Thus, *DB*-$SnCH_3$@*h*-BN is a robust QSHI with a gap of 127.19 meV, larger than the one of *DB*-$SnCH_3$@SiC by about 32 meV.

Experimentally, the stanene has been successfully fabricated by MBE on a $Bi_2Te_3$ (111) substrate,[21, 49] which demonstrates the existence of stanene film. Thus, we construct the epitaxial *DB*-$SnCH_3$@2×2$Bi_2Te_3$ HTSs by putting a single layer *DB*-$SnCH_3$ on a $Bi_2Te_3$ slab of 1 ~ 4 quintuple layers (QLs), see Fig. 6(a). The lattice mismatch between two layers is 1.67 % with the experimental lattice constant of 4.25 Å for $Bi_2Te_3$. [21] The binding energy is -1.50 eV/atom with a interlayer distance of 2.70 Å, suggesting a strong electron transfer from the $Bi_2Te_3$ substrate to *DB*-$SnCH_3$. In this case, the orbital hybridization between Te and Sn atoms is rather strong, resulting in a metallic state, as shown in Fig. 6(b). To decrease the interaction between two layers, we saturate the surface Te with hydrogen atoms, see Fig. 6(c). As is expected, it becomes a vdW HTS, which is confirmed by a large interlayer distance (3.74 Å) and a small formation energy (-0.215 eV/atom). Fig. 6(d) displays the band structure with SOC, showing a transform from metal to QSHI. The topological band gap is enhanced to 174.5 meV due to its proximity to $Bi_2Te_3$ sheet, as compared with free-standing *DB*-$SnCH_3$. Here we must point out that, though the substrate breaks the inversion symmetry of *DB*-$SnCH_3$, the substrate-induced spin-splitting of valence band is rather small. The reason can be attributed to the relatively weak interaction between two layers because of the saturation of Sn (2) atoms. These demonstrate that passivation would provide an efficient tool for screening films for their viability in the presence of substrate effects.

In conclusion, based on first-principles calculations, we predict that *DB*-$SnCH_3$ is stable and behaves as a 2D TI, with a bulk-gap as large as 148 meV, which can be further tuned by external strain. The origin of QSHE is demonstrated by $s$-$p_{x,y}$ band inversion, topological invariant $Z_2$, and helical gapless edge within bulk band-gap. We also propose SiC, *h*-BN, and $Bi_2Te_3$ as appropriate substrates to support *DB*-$SnCH_3$ to realize QSHE, indicating the high possibility for room-temperature QSHE in spintronics.

**Acknowledgments:** This work was supported by the National Natural Science Foundation of China (Grant No.11434006, No.11274143, 61571210, and 11304121), and Research Fund for the Doctoral Program of University of Jinan (Grant No. XBS1433).

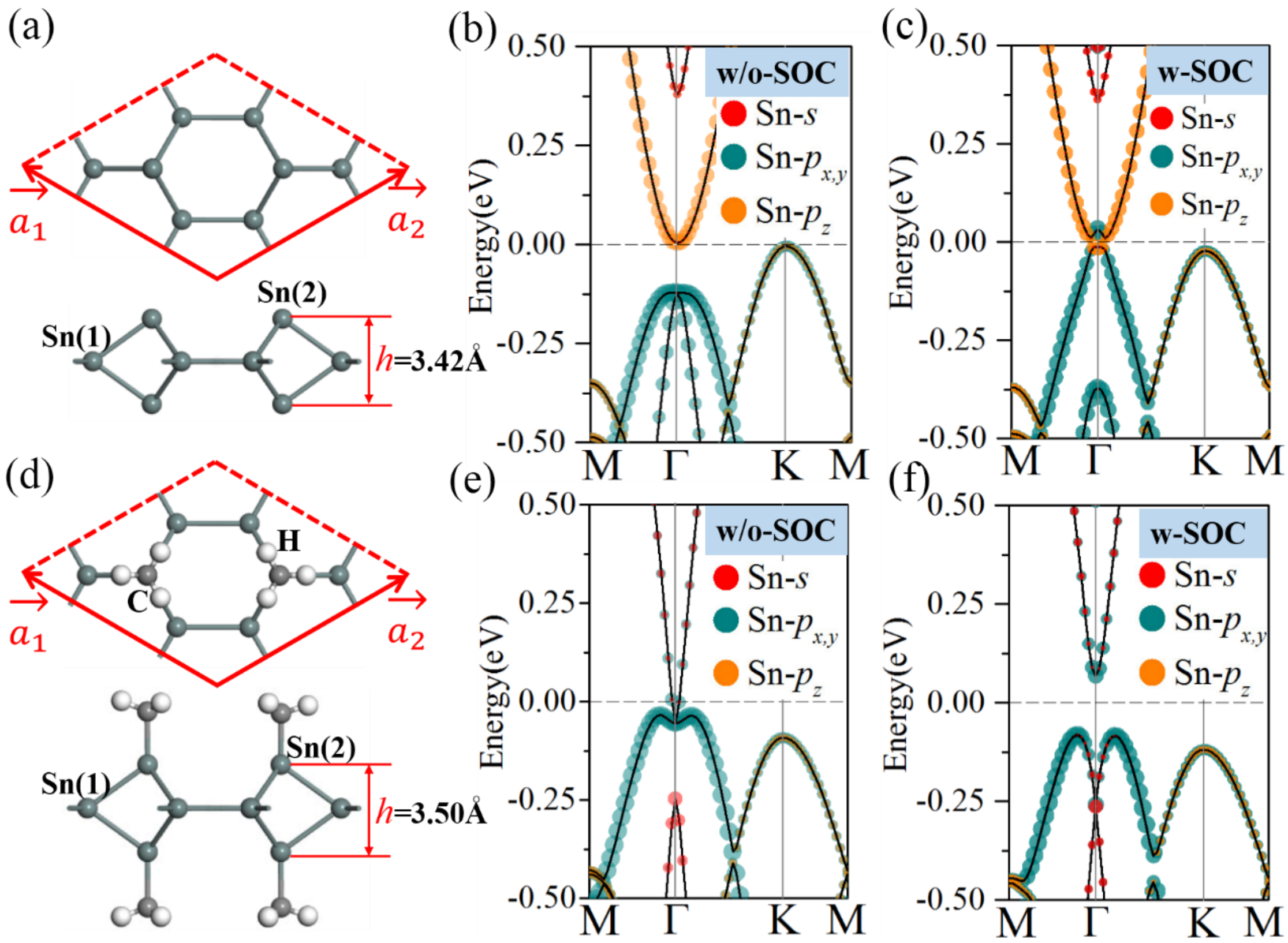

**Fig. 1** The top and side views of dumbbell stanene (a) and methyl-functionalized dumbbell stanene (d). Their band structures without SOC are shown in (b) and (e), while the ones with SOC are painted in (c) and (f). The red, green and orange dots represent the s, $p_{x,y}$, z orbitals, respectively.

(a)

| $\Gamma_i$ | Parity of $\xi_{2n}$ of occupied bands | $\delta_i$ |
|---|---|---|
| (0.0, 0.0) | +--++----++++--++++---+-+---++---+ | + |
| (0.5, 0.0) | +--++--+-++--++--++--++-+--++--++- | - |
| (0.0, 0.5) | +--++--+-++--++--++--++-+--++--++- | - |
| (0.5, 0.5) | -++--++-+--++--++--++--+-++--++--+ | - |

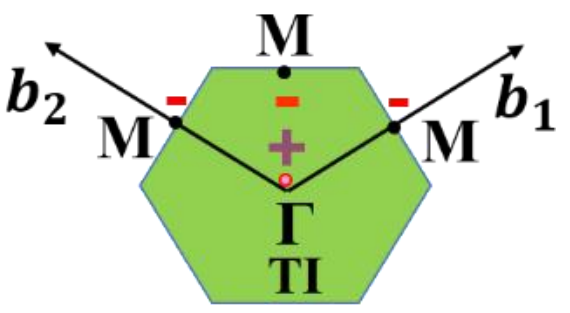

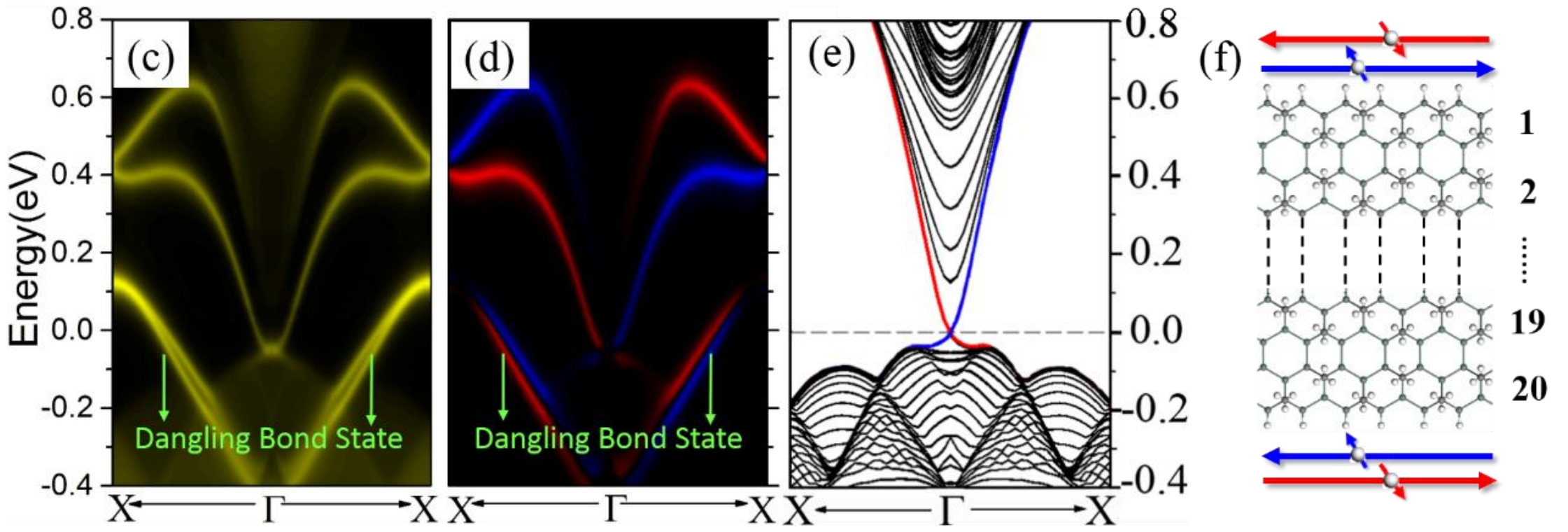

**Fig. 2** (a) Parities of the occupied spin-degenerate bands at RIM Points for *DB*-$SnCH_3$. Here, we show the parities of 34 occupied spin-degenerate bands for *DB*-$SnCH_3$. Positive and negative signs denote even and odd parities, respectively. (b) The invariants *v* can be derived from the parities of wave function at the four TRIM points $K_i$, namely one Γ point and three M points in the Brillouin zone. (c) and (d) display the electronic structures of helical edge states of *DB*-$SnCH_3$, and corresponding spin polarization in two channels. The geometry structure (f) and band structure (e) of hydrogenated nanoribbion contained 20 primitive cells are also displayed.

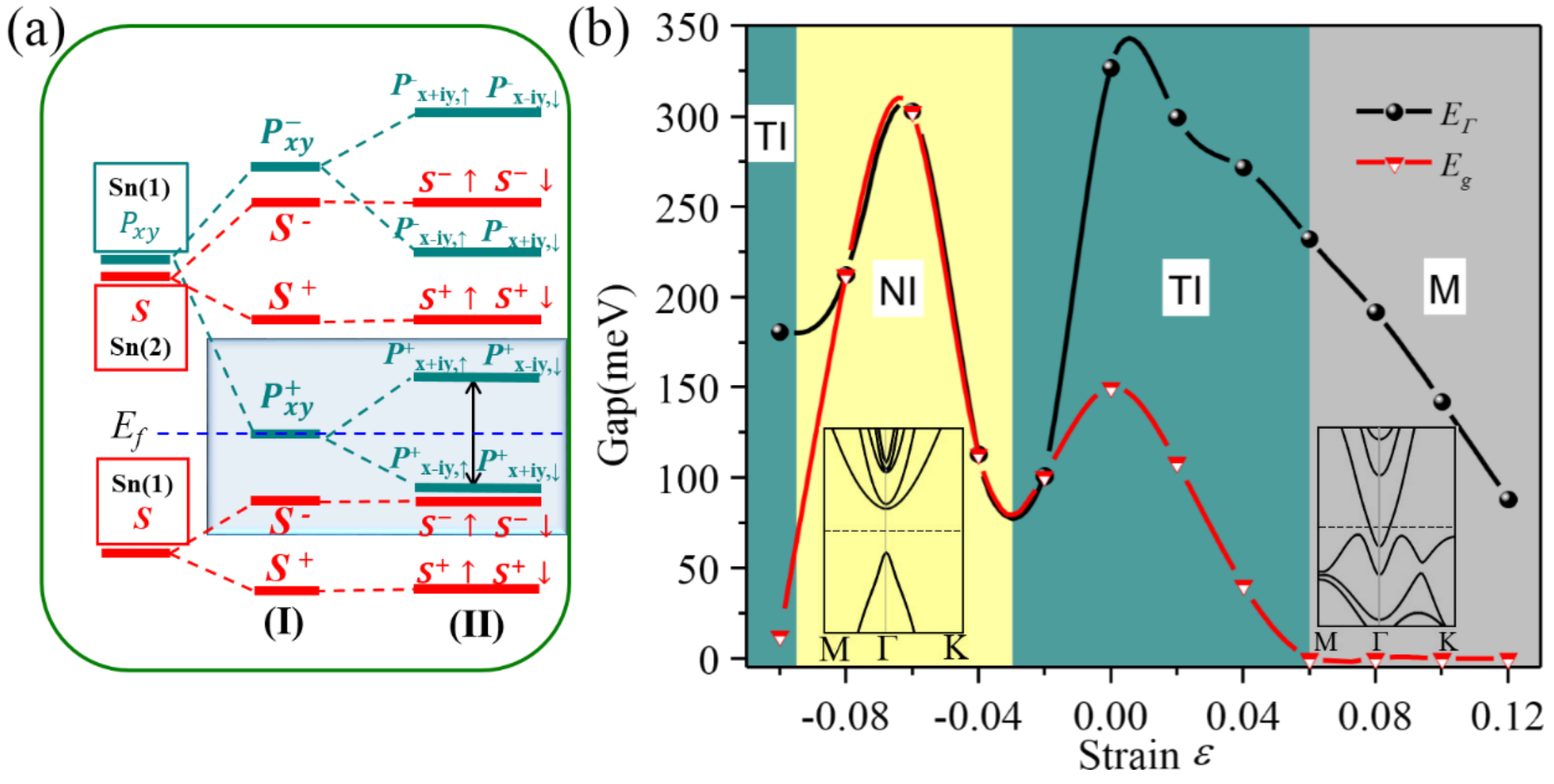

**Fig. 3** (a) The band evolution at Γ point for *DB*-SnCH$_3$ and (b) variations of band gap with respect to external strain, the inserts are band structures of normal insulator (NI) and metal (M).

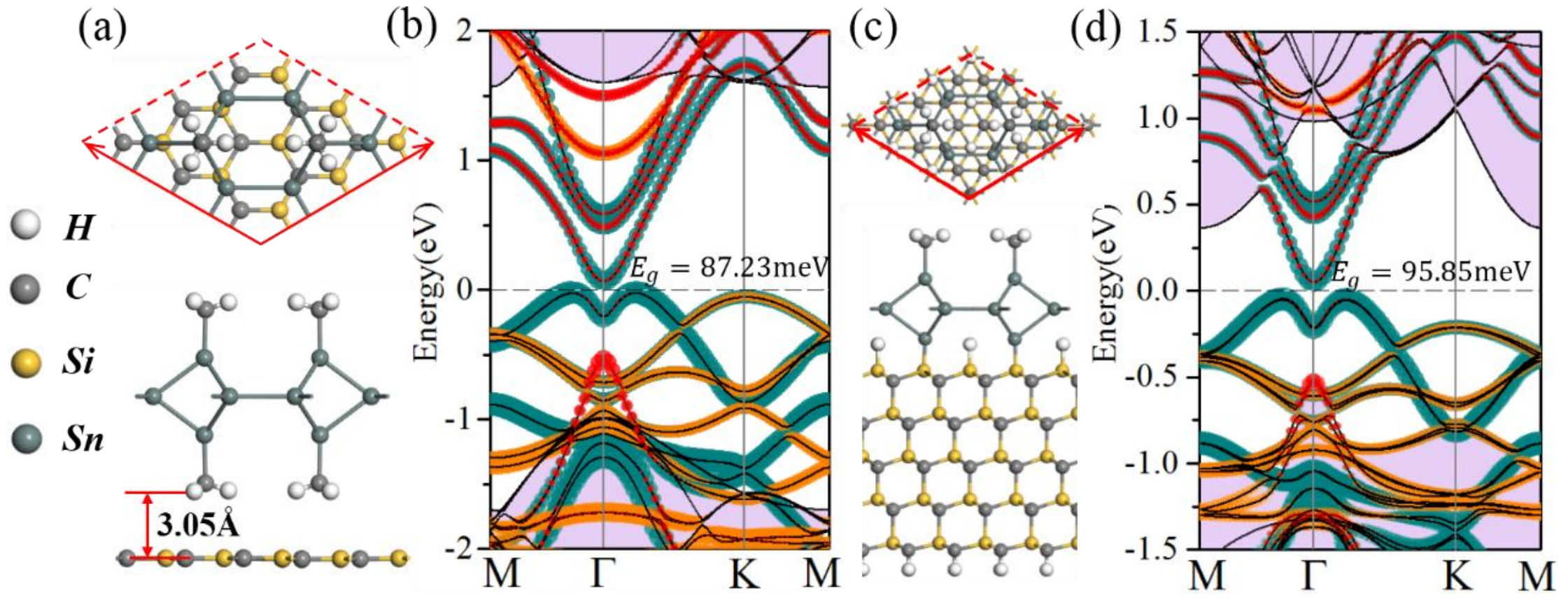

**Fig. 4** (a) (c) The top and side views of *DB*-SnCH$_3$@3×3-ML-SiC HTS and *DB*-SnCH$_3$@5-ML-SiC HTS and (b) (d) corresponding band structures.

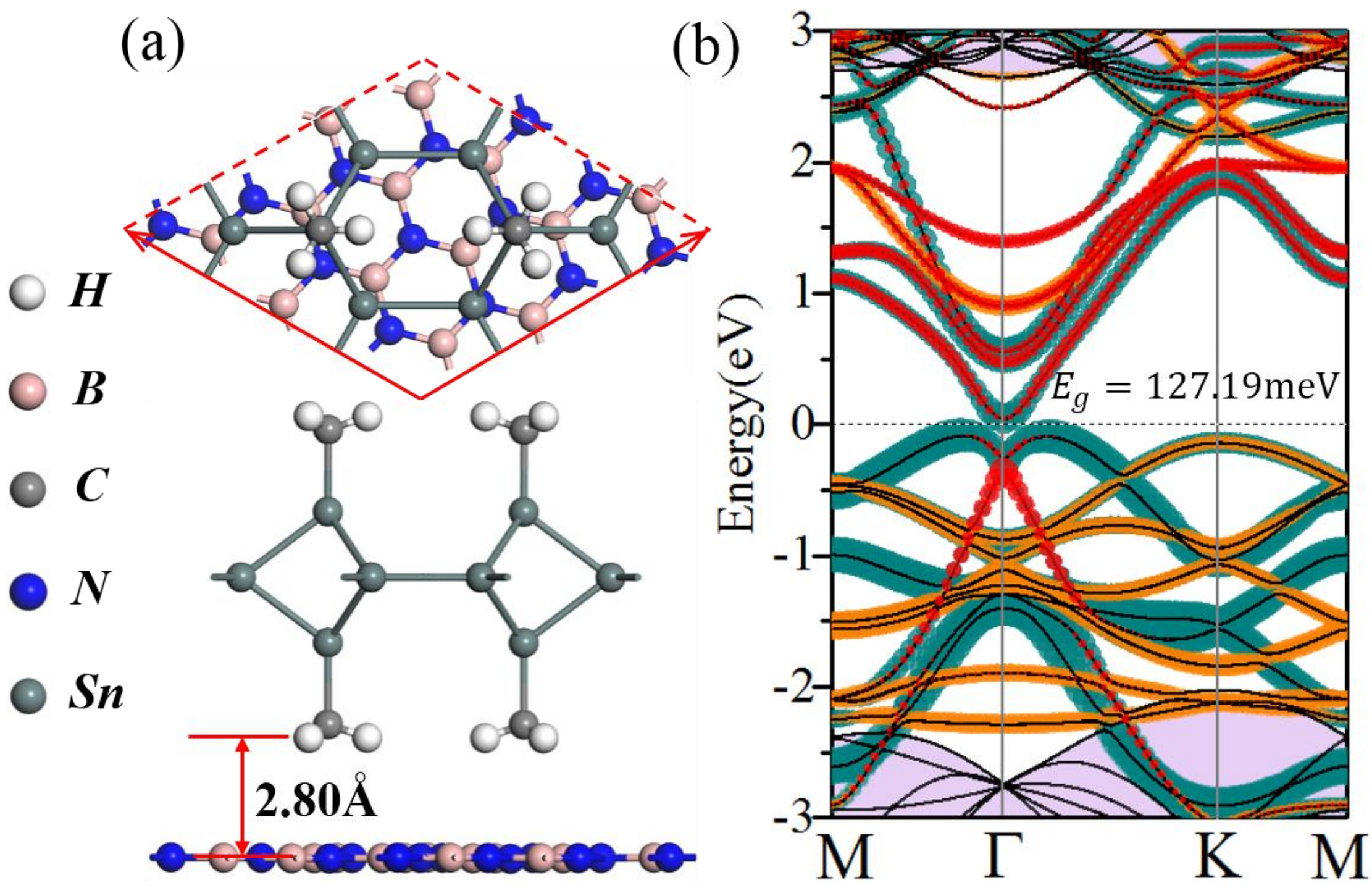

**Fig. 5** (a) The top and side views of *DB*-$SnCH_3@\sqrt{13}\times\sqrt{13}$ *h*-BN HTS and (b) corresponding band structures.

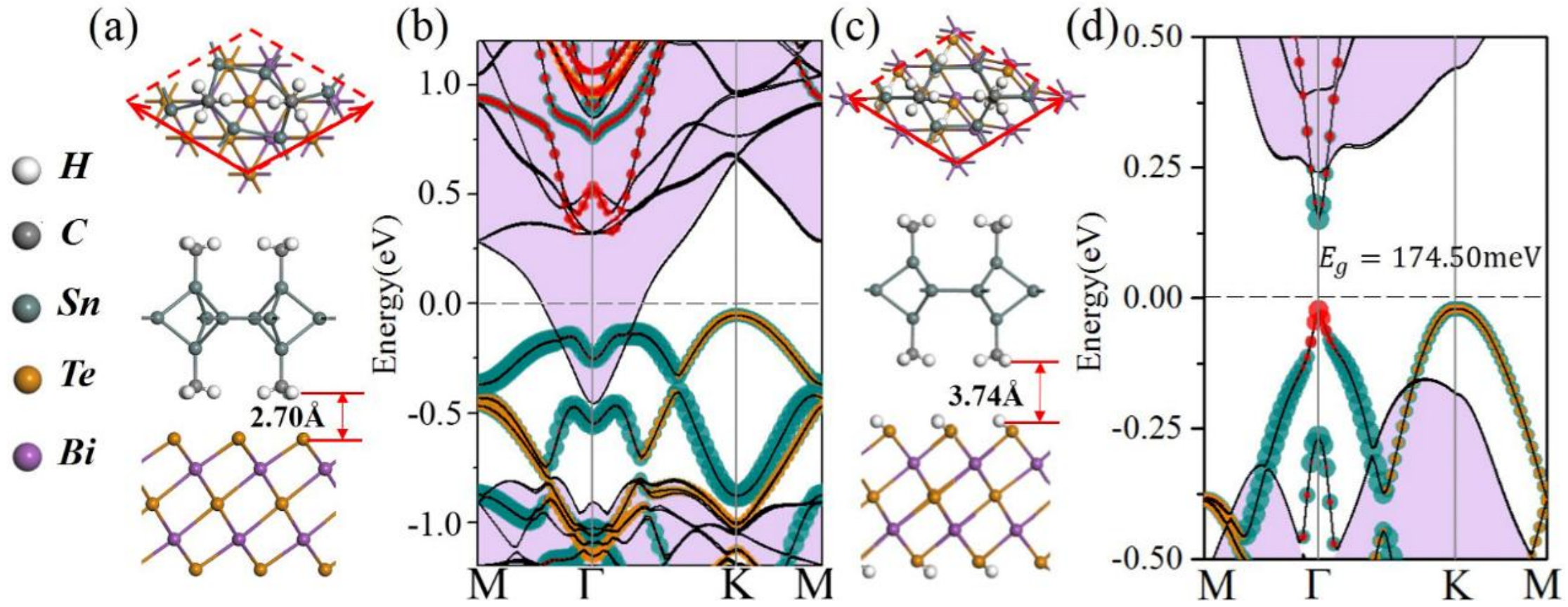

**Fig. 6** (a) and (c) The top and side views of *DB*-$SnCH_3/2\times2Bi_2Te_3$ HTSs and substrate saturated by hydrogen. (b) and (d) display the corresponding band structures for (a) and (c), respectively.